\documentclass{article}

\usepackage[margin=0.5in]{geometry}
\usepackage{graphicx}
\usepackage{hyperref}
\usepackage{subcaption}
\usepackage{float}
\usepackage{rotating}
\usepackage{amsmath}
\usepackage{placeins}

\title{\textbf{DEFENDER}: \textbf{DE}tecting and \textbf{F}orecasting \textbf{E}pidemics
	using \textbf{N}ovel \textbf{D}ata-analytics for \textbf{E}nhanced \textbf{R}esponse \\ }

\date{}
\author{Donal Simmie \qquad Nicholas Thapen \qquad Chris Hankin\\
Institute for Security Science and Technology, Imperial College London}

\begin{document}

\maketitle

\abstract{In recent years social and news media have increasingly been used to 	
		explain patterns in disease activity and progression. Social media 
		data, 
		principally	from the Twitter network, has been shown to correlate well 
		with 
		official disease case counts. 
		This fact has been exploited to provide advance
		warning of outbreak detection, tracking of disease levels and the 
		ability 
		to predict the likelihood of individuals developing symptoms. 
		
		In this paper we introduce DEFENDER, a software system that integrates 
		data from social and news media and incorporates algorithms for outbreak
		detection, situational awareness, syndromic case tracking and 
		forecasting.
		As part of this system we have developed a technique for creating a 
		location network for any country or region based purely on Twitter 
		data. We also present a disease count tracking approach which leverages 
		counts from multiple symptoms, which was	found to improve the 
		tracking 
		of diseases by 37 percent over a model that used only previous case 
		data. Finally we attempt to forecast future levels of symptom activity 
		based on observed user movement on Twitter, finding a moderate gain of 
		5 percent over a time series forecasting model. }

\section{Introduction}

The recent increases in global travel and the interconnected nature of modern
life has led to an increased focus on the threat of diseases, both established
and newly emerging. Public health officials need timely and accurate information
on disease outbreaks in order to put measures in place to contain them.
Traditional disease surveillance techniques such as reporting from clinicians 
take 1-2
weeks to collate and distribute, so finding more timely sources of information
is a current priority.

In recent years social media, especially Twitter data, has been used
to positive effect for: disease tracking (predicting the current level of illness
from previous clinical data and current social data)
\cite{collier11, culotta10, lamb13, lampos10tracking, lampos10flu}, outbreak
detection \cite{aramaki11, bodnar13, diaz12, li13} and predicting the likelihood
of individuals becoming ill \cite{sadilek12modeling}. News media has also been
used to give early warning of increased disease activity before official sources
have reported \cite{brownstein08}.

This paper presents a software system, DEFENDER, which leverages social media to
provide a combined disease outbreak detection and situational awareness
capability. Much of the previous research in these areas focuses on specific 
conditions,
with influenza being the most studied. Our approach seeks to generalise by
focusing on symptoms of disease rather than diseases themselves. A limited range
of symptoms characterises many common diseases, so a shift to symptoms adds
flexibility without a great deal of additional complexity. Specific conditions
can then be tracked by examining combinations of their symptoms. Currently,
identifying the causal explanation for a detected event is generally a manual
process. Our situational awareness system uses frequency statistics and cosine
similarity based measures, outlined in detail in a previous paper by the 
authors \cite{thapen15}, to produce
terms characterising the event and then retrieve relevant news and
representative tweets. Our generalised symptom focus extends to disease
tracking. Here we use dynamic regression to fit observed symptom levels in
social media to actual clinical disease count data. We attempt to forecast 
future values of social media
symptom data using another dynamic regression model, this time using current
count data and expected movement of symptomatic individuals as the additional
regressor.

The novel contributions in this paper are:
\begin{itemize}
\item A technique to create a data driven location network derived from social media content.
\item A software system with a general symptom focus combining event detection, 
situational awareness, disease tracking and forecasting.
\item A generalised disease tracking approach which leverages counts from 
multiple symptoms.
\item Forecasting future symptom count levels using observed people movement 
from social media.
\end{itemize}

The current state-of-the-art is discussed in section
\ref{sec:rel-work}. The data acquisition method is detailed in section
\ref{sec:data}. Our methodology is outlined in section \ref{sec:method}, and the
evaluation is contained in section \ref{sec:results}.  Section
\ref{sec:conclusion} contains concluding remarks and a discussion of areas for
further work.

\section{Related Work}
\label{sec:rel-work}

\textit{The term syndromic surveillance refers to methods relying on detection
of clinical case features that are discernible before confirmed diagnoses are
made. In particular, prior to the laboratory confirmation of an infectious
disease, ill persons may exhibit behavioural patterns, symptoms, signs, or
laboratory findings that can be tracked through a variety of data sources.}
\cite{mandl04}
\newline

The development of Google Flu Trends as a method of harnessing Internet-scale
data for syndromic surveillance \cite{ginsberg09}  has led to an increasing
focus on the Internet and social media as a means to obtain early warning of
disease outbreaks. In that research a logistic regression was performed on the
web search terms that best matched data from the US Centers For Disease Control 
(CDC) Influenza Like Illness
(ILI) data. The best performing collection of terms, in terms of fit to the CDC,
was used as the model for predicting current ILI counts. The correlations
between the regression fit and actual CDC ILI figures were high, with mean
correlations over 0.9 for all cases. However the evaluation was small and the
results have been found to be less promising than reported. According to
\cite{lazer14} the early version of Google Flu Trends was ``part flu detector,
part winter detector''.

Lampos \textit{et al.} \cite{lampos10tracking, lampos10flu} and Culotta
\cite{culotta10} use similar methods to the Flu Trends work but apply them to
Twitter data. Both studies use keyword matching to find tweets that contain flu
related terms, and both find high correlations with ground truth clinical data.
Culotta also includes a spurious match document classifier that aims to reduce
the amount of tweets that are not reporting illness, for example removing
``Bieber fever''.
Lamb, Paul and Drezde \cite{lamb13} also develop classifiers to identify health
related tweets. In their case they go one step further and separate the health
tweets into those that show signs of infection and those that merely discuss the
illness in question.

Another related area is that of outbreak detection. There are established
algorithms for alerting on disease outbreaks, such as the Early
Aberration Reporting System (EARS) developed by the CDC, which are normally
applied to actual illness case data. Some studies have attempted to apply these
algorithms to novel data sources such as social media.

One such study was Diaz-Aviles et. al \cite{diaz12}, which took a case study of 
a
non-seasonal disease outbreak of Enterohemorrhagic Escherichia coli (EHEC) in
Germany. Their study applied EARS to Twitter data. They searched for tweets from
Germany matching the keyword ``EHEC'', and used the daily tweet counts as input 
to
their epidemic detection algorithms.
Using this methodology an alert for the EHEC outbreak was triggered before
standard alerting procedures would have kicked in.

One of the most interesting recent studies in this area was carried out by Li \&
Cardie \cite{li13}. Using Twitter data they built a Markov Network able to
determine when the flu levels in a US state had reached a ``breakout'' stage 
when
an epidemic was imminent. Their system takes into account spatial information by
building a simplified network map of the US, where each state is a node
connected to its neighbouring states. It also takes into account Twitter's daily
effect - the fluctuation in the number of tweets posted based on the day of the
week.

Forecasting the future spread and impact of disease epidemics is hugely
important for public health, and epidemiologists have developed various models
to allow this, the most common being the SIR (Susceptible, Infected, Recovered)
model. When applied to a known population and a known disease these models can
accurately predict the epidemic curve, which describes how a disease affects a
population over time. These models require information on the characteristics of
the disease which is missing in the case of syndromic surveillance using social
media. The primary work on developing novel methods of forecasting from social
media data has been carried out by Sadilek
\textit{et al.}

They first worked on tracking the spread of disease in a social network and via
personal interaction \cite{sadilek12modeling}. This study was based on a sample
of geo-tagged tweets originating in New York City. They developed a
machine learning classifier to identify individuals exhibiting symptoms of
illness.
In the process they identified that approximately 1/1000 tweets concerned self
reporting of illness. Once they had identified symptomatic individuals they 
were able
to demonstrate that Twitter users who were friends of the affected users were
more likely to subsequently become ill than randomly selected users, and that
users who had tweeted at the same time and place as the symptomatic individuals 
were
also more likely to subsequently become ill.

In a further paper \cite{sadilek2012predicting} the authors built on this work
to develop a model allowing them to predict the future level of influenza in US
cities by modelling travel patterns of Twitter users. They collected geo-tagged
tweets from the 100 busiest commercial airports in the US, and used their
classifier to identify symptomatic individuals. They then built up a model of 
the
travel patterns between airports by identifying users who had tweeted from
multiple locations on subsequent days. Using this data they found that the most
important factor in predicting the prevalence of flu in a given city was the
number of symptomatic passengers that had flown into the city over the previous 
seven
days.

\section{Data}
\label{sec:data}

Data for our work was obtained from Twitter's live streaming API using a 
geographical
bounding box containing England and Wales. News data was collected from 14
national and regional news sources, using a daily RSS download. For evaluation
purposes we used clinical data from Public Health England in the form of the GP
In Hours Weekly Bulletin \cite{gp-in-hours-2014}.
The data collection period was from February to August 2014. A total of
84,438,013 tweets and 12,130 news articles were collected during this period.
Retrieving tweets in this manner only returns those which contain exact
geo-location information, which form around 1.6\% of the total number of tweets 
\cite{leetaru13}.

\subsection{Symptom Focus}
\label{sec:symptom-focus}

The system was developed with a focus on symptoms of illness.
This approach was adopted since a limited range of symptoms characterise many common
diseases, and the identification of disease from the symptoms presenting is itself a complex issue.

The Freebase online database \cite{freebase} was used to capture a
representative set of symptoms. The exact process is described in
\cite{thapen15}. The end product was a list of 46 symptoms, each represented by 
groups of keywords.
The number of tweets matching symptom keywords was tracked on a daily basis for 
each geographical area being monitored.

\subsection{Noise Removal}
\label{sec:noise-remove}

\begin{table}[H]
	\centering
	\begin{tabular}{l}
		\hline
		``yet again an instance of fear over haemorrhaging of labour votes. ukip ukip.''
		\\
		``my 2014 has so far been sponsored by the common cold \#spons"\\
		``sore head, sore eyes, sore legs and a runny nose. looking gooooood!"\\
		``lets not make rash statements!"\\
		``this game boils my blood !!!! \#flappybirdsshoulddie''\\
		``i hate hay fever -\_-"\\
		\hline
	\end{tabular}
	\caption{\textbf{Key-term matching tweet examples:} sampled from the
		health classifier test set.}
	\label{tab:keyword-matches}
\end{table}

The primary problem with using unstructured social media data is that it is very noisy. 
Tweet content matching a keyword may not relate to illness at all, but be an unrelated use of the 
word or general discussion of illness rather than reporting (see Table 
\ref{tab:keyword-matches} for example tweets).

In order to overcome the noise issue we developed two classifiers to identify health-related content. The
first is a semi-supervised Support Vector Machine (SVM) implementation that
classifies tweets as being health related or not. The second is a Naive Bayes
classifier that performs the same task for news articles.

The approach that has been used for tweet classification is an adapted version
of the model used by Sadilek \textit{et al.} \cite{sadilek2012predicting}. This
classifier model uses a cascading approach where an initial set of manually
labelled tweets is expanded twice to form more training data for subsequent
supervised classifier to use. The \textit{LibShortText} \cite{libshorttext} SVM
library was used to perform documentation classification. 

The set of manually labelled tweets, 4600 in total, was sampled from the
tweet datastore that consisted of over 30 million tweets at the time of
collection. We performed a structured sample, limiting the number of each
symptom group sampled from the population set. This allowed important but infrequent terms
like tonsillitis or chest pain to have more instances in the sample set and
boosted classification accuracy.  The final classifier achieved a classification
accuracy of 96.1\% on a test set of manually classified tweets.

We also developed a classifier to determine which news articles were health 
related. This utilises a more traditional supervised approach
where training data is created by segmenting articles definitely from health
only sources and those from news sources that are not exclusively health feeds.
This training data was then fed into a standard Naive Bayes text classification
algorithm using unigrams as features. This classifier was found to have an 
accuracy of 84\% on a a test set of manually classified articles.

\section{Methodology}
\label{sec:method}
\subsection{Location}

In a previous paper \cite{thapen15} we developed a methodology for identifying
areas of high tweet activity within a country or region. This was done by
clustering the geo-located tweets using the DBSCAN algorithm. When applied to
the UK the resulting areas were found to correlate with the major towns and
cities.

We extend this work by considering each area identified as a node within a
complete undirected graph network. Edge weights between nodes are set as
observed people movement between nodes.
This is calculated from our data by counting the number of people tweeting from
different nodes on the same day (midnight to midnight), averaged over the entire
period of the study.

There are multiple advantages to this data-driven generalised location system.
Firstly as mentioned in \cite{thapen15} it is transferable to any new area 
within which
there is sufficient tweet density to create clusters. Secondly locations which
are not spatial neighbours but exhibit high population movement, for example
London and Birmingham, can be considered to be ``closer'' than they are by
distance alone. This is advantageous considering that human contact is the most
important factor in spreading infectious disease \cite{clayton93} and thus
likelihood of movement between two areas increases contact exposure chance.
Finally it allows for the leveraging of existing graph based algorithms and
techniques devised for analysing the structural properties of graphs such as
PageRank and betweenness centrality. 

The main disadvantage of this generalised approach is that mapping back to a
region's own hierarchical administrative system (for ground truth evaluation of 
system accuracy) requires some effort. 

\begin{figure}[H]
        \centering
        \includegraphics[scale=0.3]{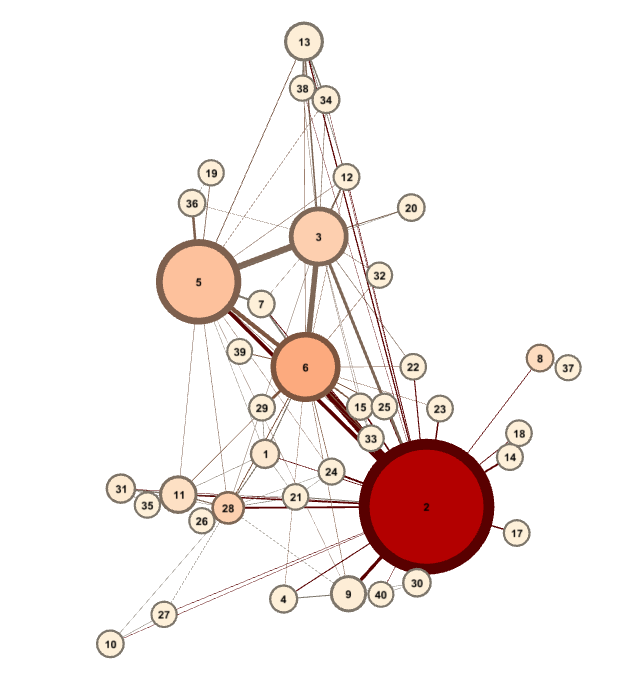}
        \caption{\textbf{The DEFENDER Location Network for England and Wales:} 
        Nodes are coloured 
        according to their calculated PageRank within the network}
        \label{fig:defender-network}
\end{figure}
	
\subsection{Architecture}	

\begin{figure}[H]
        \centering
        \includegraphics[scale=0.7]{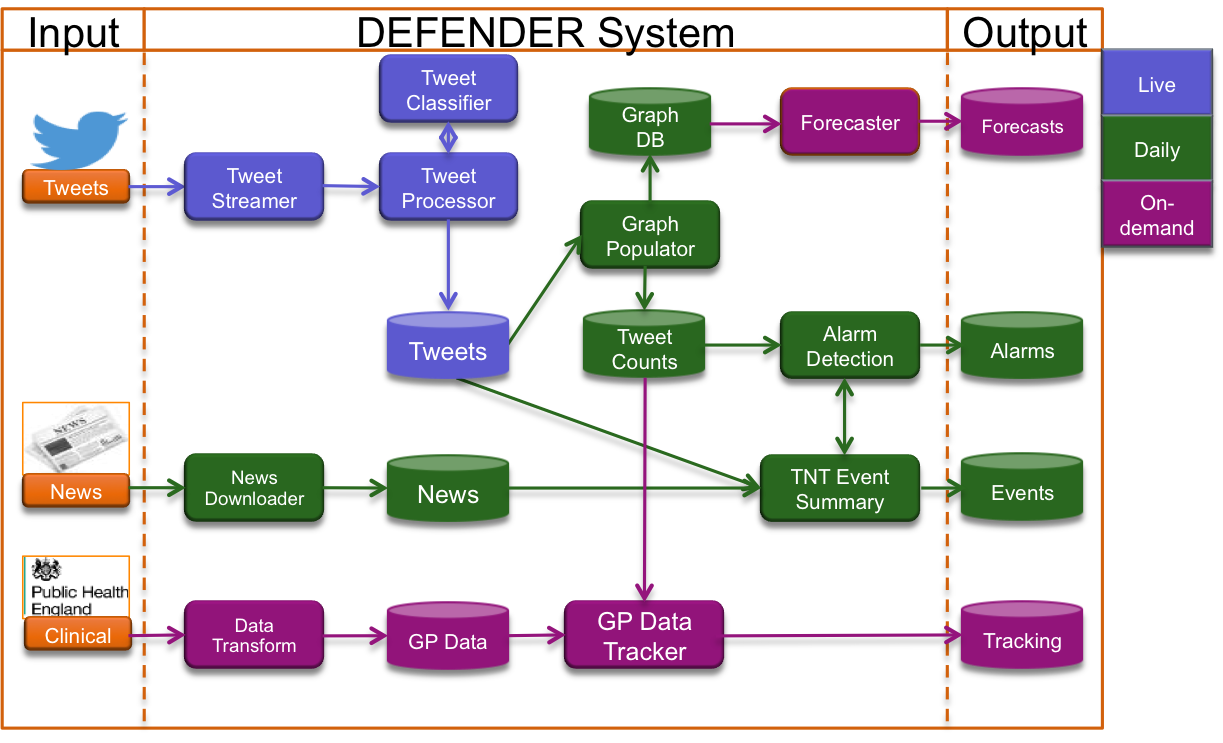}
        \caption{\textbf{Architecture of DEFENDER:} the left area shows inputs
        to the system: tweets, news articles and clinical data. The middle
        section details the internal services and data stores and the area on
        the right-hand side highlights the outputs which are available to be
        viewed or queried against. The different colours denote the data
        processing category.}
        \label{fig:defender-arc}
\end{figure}

The DEFENDER data processing pipeline deals with data at three different 
temporal resolutions:
live, daily and on-demand. Live is data that is collected and processed within
seconds of its conception. Daily data processing tasks are scheduled to
execute once a day. On-demand data is produced in response to a request from a
system user. The system architecture is delineated in Figure
\ref{fig:defender-arc}. 

Tweets are the primary data sources and are ingested into the system as they are
posted to Twitter. The tweets are classified to ensure they are health related.
Additional pre-processing then occurs where tweets are assigned to nodes in the
network; named entities and symptom keyword matches are also extracted at this
point. The graph populator service runs daily against the dataset of classified
processed tweets. It extracts the symptom time series counts and creates a graph
where the nodes (locations) have daily symptom count properties and the edges
are the observed people movement between node pairs.

The graph database is used as an input by the forecasting service. The observed
people movement between nodes and the amount of people exhibiting symptoms
within an area provide enough information to estimate the number of symptomatic 
people
moving into an area from its neighbours. The node/symptom tweet counts are used
as input for both outbreak detection and the clinical disease tracking
application. The TNT event summarisation service (see \cite{thapen15}) uses the
alarms generated by the outbreak detection algorithm and fetches the source
tweets from the processed tweet store with the aim of revealing any potential
cause of the symptom event. The news data, which is ingested daily into the
system, is also used as input to the event summarisation service. News data is
linked by finding terms that produce good matches from searches of news article
stores. The linked events (tweets and news) are stored in an event database
which can be viewed via a front-end or queried directly. The disease tracking
application attempts to forecast current actual levels of illness from
lagged (one week) clinical source data. The past GP case visits and current 
tweet
count data are used to achieve this. 

\subsection{Services}

\subsubsection{Early Warning Detection}
\label{sec:ewd-method}

Early warning of disease outbreaks permits health officials to develop timely
intervention strategies and can prevent large scale crises. The DEFENDER system
uses an Early Warning Detection (EWD) methodology which was described in a
previous work by the authors \cite{thapen15}. This uses the EARS
syndromic surveillance algorithm to trigger alarms based on spikes in 
symptomatic
tweet activity. By applying customised filtering criteria, including removing
those alarms with a low deviation from the time series median, we are able to
produce robust high confidence alarms, as evaluated in \cite{thapen15}.

\subsubsection{Situational Awareness}	
\label{sec:sa-method}

\begin{figure}[H]
        \centering
        \includegraphics[scale=0.3]{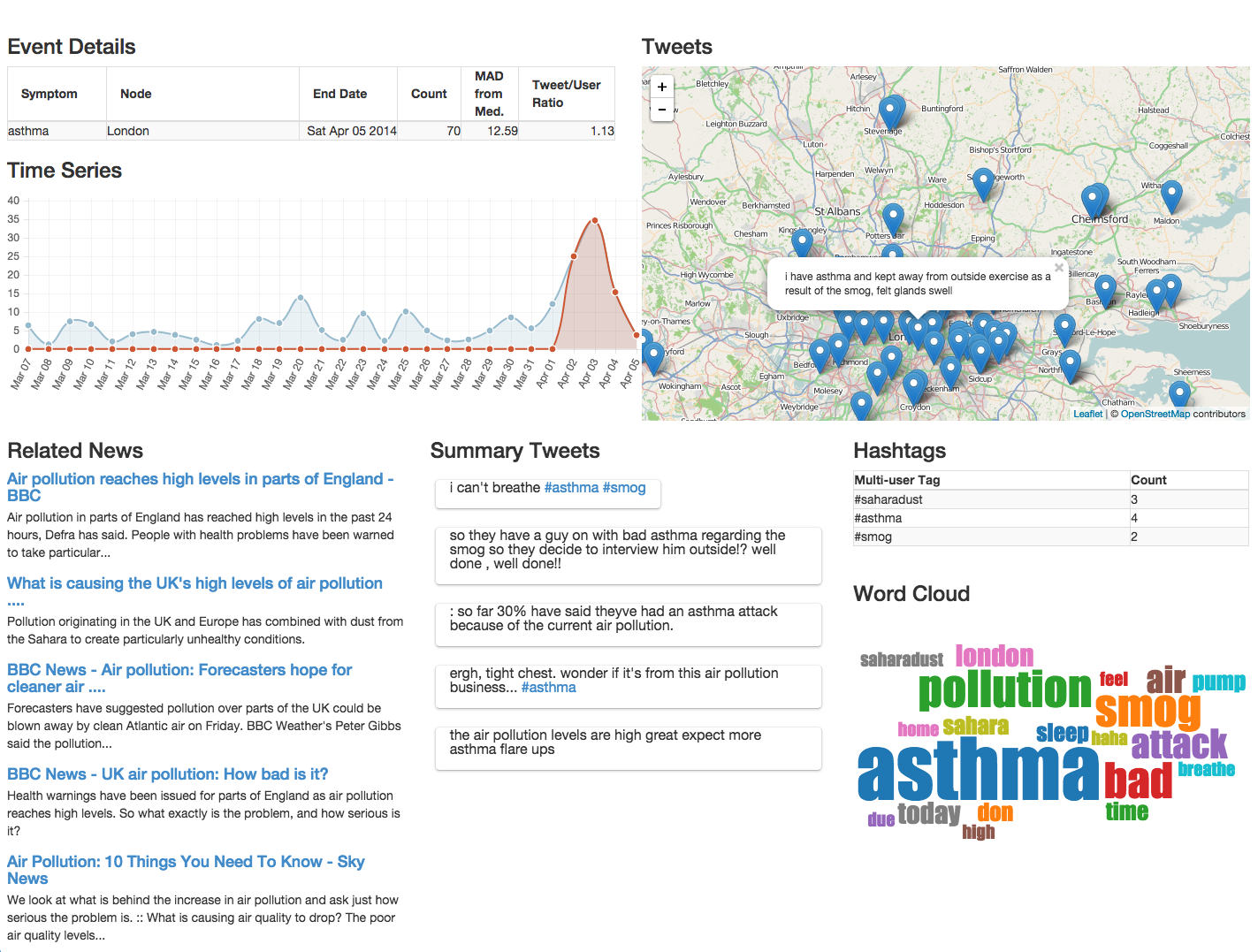}
        \caption{\textbf{DEFENDER Situational Awareness screen showing asthma 
        event in London from April 2014.} Moving counter-clockwise from the top 
        left the screen shows: \textbf{1. Event Details:} Basic information 
        about the timing and severity of the event. \textbf{2. Time Series:} A 
        graph showing the symptom counts for the previous 30 days with the 
        alarm period highlighted in red. \textbf{3. Related News:} Most 
        relevant news articles selected by TNT algorithm. \textbf{4. Summary 
        Tweets:} Most relevant tweets selected by TNT. \textbf{5. Word Cloud:} 
        Visual representation of words used most frequently in the tweets. 
        \textbf{6. Hashtags:} Tags used by multiple users within the event. 
        \textbf{7. Tweets:} Map showing locations of individual tweets. }
        \label{fig:detail-view}
\end{figure}

The situational awareness module provides additional context to a reported
health outbreak alarm. This could be used by a public health official to
determine the cause, location and importance of the alarm.
We use the tweets associated with an alarm (those that match the alarm keyword
and originate from the same location and time) in order to generate the
situational awareness report. We use the TNT algorithm, developed for this
system and described in \cite{thapen15}, to extract terms which are specific to
the event rather than the baseline of symptomatic tweets from the area. These
terms potentially describe the event and can be used to retrieve relevant news
articles and rank the tweets which best summarise the event. Additionally we
also extract hashtags used by more than one user, frequent terms and
geo-coordinates from the alarm tweet set.
This event metadata is stored as part of the event along with the tweets from
which they were extracted. When relevant news is found the article metadata and
text is also stored as part of the linked health event. A system user can then
query the event database for symptomatic events or use a front-end to provide a
visual summary of a specific event.  We have developed a front-end using open
source web technologies as an example visualisation platform. An illustrative
example of a detected event is presented in Figure \ref{fig:detail-view}.

\subsubsection{Tracking}
\label{sec:tracking-method}

The tracking module aims to predict current case count data from two
time series. The first is historical case data available until the previous
time point, the second is tweet count data available up to and including the
current time point The assumption is that adding the current tweet count data
series will aid in predicting the current case count data.

The forecasting technique employed by this algorithm is dynamic regression,
specifically regression with ARIMA (Autoregressive Integrated Moving Average)
errors. This combines the time series focused dynamics of ARIMA models with the
potential for additional regressors to explain a certain amount of variance in
the data.

There are two modes of operation for the disease tracker: 1) model training
(including feature selection) and 2) symptom tracking using trained models. The
model training algorithm uses a candidate set of symptom features (terms)
provided as input for the case data in question. For example flu cases may use:
sore throat, fever, cough, flu, headache or other appropriate terms. From these
terms all possible combinations of these terms (up to 4 terms at once) are
used. The best performing keyword combinations, measured by lowest Mean Absolute
Error (MAE) on a cross-validation window of 4 time periods (28 days training 7
days test) are persisted as the best model for that symptom/node combination. 

An interesting note is that originally the tracker used the AICC criterion
\footnote{http://en.wikipedia.org/wiki/Akaike\_information\_criterion\#AICC} to
determine the best forecast fit to the trained data and select forecast models
based on this. However this was found to overfit the data and have weak
predictive power for unseen data. Choosing the best fits from the MAE of
4-fold cross-validation on training data improved the accuracy on unseen data
considerably.

This is a lengthy process as it involves fitting models for all possible
combination of choosing 1,2,3,4 terms from all candidate terms. For
example, for ILI cases in the evaluation 14 flu terms are used; the total 
number 
of model fits is then: $^14C_1 +
^14C_2 + ^14C_3 + ^14C_4 = 1470$. This however need only be done periodically.
Once the best model for a GP case/location pair has been ascertained that
feature set can be used to track new data in near real-time.

The tracking evaluation section \ref{sec:tracking-eval} details the results of
an experiment to determine the effectiveness of this approach and compares the
chosen model regression with ARIMA errors (case series and tweet count series)
to alternatives models including: regular ARIMA time series forecasting, a
random walk forward and a naive forecast that projects forward the mean of the
existing series.

\subsubsection{Forecasting}
\label{sec:forecast-method}

The forecasting system aims to predict the future level of tweet symptom
activity across the node network. Since the system is symptom
based it was not possible to employ standard disease modelling techniques such
as SIR modelling, since the details of the disease such as the transmission and 
recovery rates are unknown. The information available to the system is a time
series of the tweet count activity for each node and symptom, and also the
information about the movement of people between nodes.

The assumption that was therefore made when trying to forecast future tweet
counts was that knowing the future influx of symptomatic people from 
neighbouring
regions to an area would lead to a more accurate prediction of future tweet
count values. In order to estimate the influx of symptomatic people from node A 
to node
B the following simple equation was used:

\begin{equation}
\text{Influx} = \frac{\text{Number of symptomatic individuals in Node A}}{\text{Total number of individuals in Node A}}\times{\text{Number of people typically travelling from A to B}}
\end{equation}

In order to use this value as a regressor in a predictive model an iterative
approach was used.

Assume that the system is making a forecast of tweet counts, starting at day $t_0$. We
have information on the tweet counts and influx of symptomatic people from the 
previous
days. The model is therefore trained to use the current day's tweet count
figure and the previous day's influx figure. A prediction is then made for
$t_1$, allowing the influx figure for $t_1$ to be calculated. This influx figure
is then used in the prediction for $t_2$ and so on.

Several forecasting methods were trialled before an ARIMA model was settled on.
A Holtz-Winters exponential moving average model was ineffective due to the lack
of strong seasonality in the tweet data. Potentially this model could be 
applicable if several years of data were available, since many illnesses, in
particular influenza, exhibit an annual pattern. A machine learning
forecasting approach available in the \textit{Weka} toolkit \cite{hall2009weka} 
was also trialled,
but it was unable to deal with the strongly zero-weighted time series found for
many node/symptom combinations, tending to predict astronomically high values if
any cases appeared in the forecasting period.

 The best performing models were found to be two
commonly used forecasting algorithms. The first is an ARIMA model that uses the historical tweet counts alone to
predict future values. This is used as a baseline comparator to determine
whether the addition of movement data improves forecasting accuracy. The second
adds an additional regressor in the form of future incoming symptomatic people 
to a
node. It is hypothesised that this extra signal should confer more information 
than knowing the current
symptom tweet activity does alone. This is evaluated in section
\ref{sec:forecast-eval},

\section{Results}
\label{sec:results}
The DEFENDER system evaluation has been carried out by evaluating the core
components: early warning detection, situational awareness, disease
tracking and forecasting. The first two of these have been primarily evaluated
in our earlier work \cite{thapen15}.  The disease tracking and
future symptom count forecasting algorithms are evaluated in the subsequent
sections; both are evaluated by comparison to similar models. Additionally we 
carried out an initial analysis of our graph location network based on Twitter 
data.

\subsection{Location Network Analysis}

When studying disease outbreaks it is important to identify movement patterns, since these
can influence disease spread as well as the public health reaction.
In order to examine the importance of each node in the network with respect to 
movement we calculated its edge-weighted PageRank. We then ranked the nodes by 
PageRank and by the number of tweets assigned to them in a one month period.

Examination of both lists reveals that some nodes are more central to the
network than their population would imply. For example Bristol is ranked 8th by
population, but moves up to 5th place by PageRank, ahead of Cardiff and
Newcastle. This may be due to its location within the UK's transport network, in
which it acts as a broker between the South West, Wales and the rest of the
country. In general it can be seen that central towns and those along the
'spine' of the UK running from Birmingham up to Manchester score more highly on
PageRank, while peripheral towns score less highly. This information could be
incorporated into the early warning system by upgrading the importance of alerts
in those areas with a high PageRank, since individuals, and therefore diseases,
are more likely to travel to and from these areas.

\begin{table}[H]
	\centering
	\begin{tabular}{lrr}
		\hline
		Node & ID & Tweet Count \\
		\hline
		London & 2 & 2,625,273  \\ 
		Manchester & 5 & 1,422,641 \\ 
		Birmingham & 6 & 1,056,275  \\ 
		Leeds & 3 & 810,153  \\ 
		Newcastle & 13 & 336,429  \\ 
		\hline
	\end{tabular}
	\quad
	\begin{tabular}{lrr}
		\hline
		Node & ID & PageRank \\
		\hline
		London & 2 & 0.16370  \\ 
		Birmingham & 6 & 0.10796  \\ 
		Manchester & 5 & 0.08667 \\ 
		Leeds & 3 & 0.07985  \\ 
		Bristol & 28 & 0.03549  \\ 
		\hline
	\end{tabular}
	\caption{\textbf{Top 5 nodes by Tweet Count and PageRank.} The left table is
	sorted by observed tweet counts from a sample period. The right table is from
	the PageRank of the node within our location network. Note that these nodes do
	not correspond exactly to the urban areas for which they are named.}
	\label{tab:top-nodes-count}
\end{table}

\subsection{Tracking Evaluation}
\label{sec:tracking-eval}

The disease tracking evaluation covered a 132 day period from February 11th to the
22nd of June 2014. GP case data was retrieved for four different cases:
Influenza-like Illness (ILI), Gastroenteritis, Diarrhoea and Vomiting at the UK
local authority level from Public Health England \cite{gp-in-hours-2014}.

A data mapping was performed from tweet counts in our node network to the local
authority level. The local authority boundary areas were retrieved from the
Ordnance Survey \cite{boundary-line}. These were converted from a shapefile
format to GeoJSON. All of the symptom cases for the period were then assigned
from the local authority level to the nodes in the location system. The local authority data was provided at a weekly
frequency, therefore to give daily counts so that they could be on the same scale as the
tweet data, linear interpolation was used producing 7 day values from one weekly
figure.

Seven different forecasting models were implemented for evaluation:

\begin{itemize}
	\item ARIMA: autoregressive time series forecasting uses the historical case
	data only to predict future values.
	\item ARIMA with regression: uses the historical case data and tweet count
	information to predict future values.
	\item ARIMA with regression of simple moving average: uses historical case
	data and a simple moving average (one week smoothed value) of the tweet
	count.
	\item ARIMA with regression and weekly lagged difference between time series
	and regressor: performed due to the success of Lazer's model \cite{lazer14},
	this model uses historical case data, current tweet count data and a lagged
	(one week ago) difference between the two values.
	\item Naive Control Models:
	\begin{itemize}
		\item Mean: simple forecast projecting forward the mean of the series.
		\item RWF: random walk forward: the last value in the observation series is
		used for all forecasted values.
		\item RWF with drift: same as previous except a directional drift is
		included.

	\end{itemize}
\end{itemize}

A cross-validation testing procedure was employed rather than a single
train/test period, in case the position in the series affected the
results. Four folds were used, each consisting of 33 days of observations (26
days training data, 7 days test). The evaluation metric used was the Mean
Absolute Error, which quantifies the difference between the fitted and actual
figures. This was recorded for each node/case pair. The overall results,
presented in Table \ref{tab:mean-all-mae}, show that the model with combined
time series forecasting and tweet count regression (ARIMA Reg.) was the best 
performer. This
is in-line with the assumption that had been made that adding tweet count data 
would
aid in tracking the GP case data effectively. The regression with ARIMA errors
model performed almost 40\% better on average than the other models.
Interestingly a similar model which performed well in \cite{lazer14} was
beaten by the time series forecast. There are many possible choices for choosing
a lagged predictor so it is
left as future work to examine if any of these could improve upon these results.

\begin{table}[H]
	\centering
	\begin{tabular}{lrr}
		\hline
		Model & Mean MAE & Percent Diff. (Min) \\
		\hline
		ARIMA & 13.05 & 37.21 \\ 
		ARIMA Reg. & 8.20 & 0.00 \\ 
		ARIMA SMA Reg. & 17.24 & 52.45 \\ 
		ARIMA Lag. Reg. & 15.06 & 45.59 \\ 
		Mean & 31.50 & 73.98 \\ 
		RWF & 19.06 & 56.99 \\ 
		RWF Drift. & 22.02 & 62.78 \\ 
		\hline
	\end{tabular}
	\caption{\textbf{Mean of all MAE errors for each node/case pair}. The percentage
		difference from the minimum observed value is also presented. (4-fold 
		cross-validation)}
	\label{tab:mean-all-mae}
\end{table}

The mean of all node/case MAEs is a mean figure hence it is vulnerable to
outliers distorting the results. An alternative approach to confirm the success
of this model is to see the fraction of node/case pairs where the models have had
the lowest (i.e. best) MAE. From Table \ref{tab:fraction-mae} it is clear that 
this
model is the best performer with a significantly higher fraction than all other
models for all cases. These aggregate level results are from all node/case
pairs.

\begin{table}[ht]
	\caption{\textbf{Fraction of node/case pairs where model has minimum MAE} 
	(4-fold cross-validation)}
	\label{tab:fraction-mae}
	\centering
	\begin{tabular}{lrrrrrrr}
		\hline
		GP Cases & Arima & Arima Reg. & Arima SMA Reg. & Arima Lag. Reg. & Mean  & Rwf & Rwf Drift. \\
		\hline
		ILI & 0.05 & 0.57 & 0.14 & 0.10 & 0.05 & 0.10 & 0.00 \\
		Vomit & 0.23 & 0.59 & 0.00 & 0.05 & 0.00 & 0.09 & 0.05 \\
		Diarrhoea & 0.13 & 0.48 & 0.00 & 0.13 & 0.04 & 0.22 & 0.00 \\
		Gastroenteritis & 0.00 & 0.57 & 0.09 & 0.04 & 0.04 & 0.22 & 0.04 \\ 
		\hline
		Average (Mean) & 0.10 & 0.55 & 0.06 & 0.08 & 0.03 & 0.16 & 0.02 \\
		\hline
	\end{tabular}
\end{table}

\subsection{Forecasting Evaluation}
\label{sec:forecast-eval}

Evaluation of the forecasting module covered the period from February 11th to
June 3rd. The system was evaluated by repeatedly performing
forecasts and then checking the predictions against the actual observed tweet
counts. This cross-validation approach was very similar to that used in the
evaluation of the disease tracking module. Four folds were used, each consisting
of 28 days of training data and 7 days for testing. The folds started every 28
days commencing on February 11th, so the testing week for the previous fold was
allowed to overlap with the training period for the next fold.

Five of the most common symptom groups were selected for testing: Sore
Throat, Tonsillitis, Common Cold, Flu, Cough. In each fold, all of the above
symptoms were forecast for every node, using an ARIMA autoregressive model as a
baseline, and an ARIMA model including incoming tweet count data as the favoured
model.

In order to obtain an accuracy measure the Mean Absolute Error was again used.
For this evaluation the difference between the final forecast figure (at $t_7$)
and the actual figure was calculated at every node and an average was taken.
These errors were then again averaged over all folds in order to ensure that the
results were not skewed by especially favourable or unfavourable time periods.

The results are shown in Table \ref{tab:forecast-errors}, with the influx data
giving a 5.8\% improvement in forecasting accuracy.

\begin{table}[H]
	\caption{\textbf{Mean of all MAE errors for each symptom, showing both 
	forecasting methods}. (4-fold cross-validation)}
	\label{tab:forecast-errors}
	\centering
	\begin{tabular}{lrr}
		\hline
		Symptom & ARIMA & ARIMA with influx data \\
		\hline
		Sore Throat & 1.37 & 1.43 \\ 
		Tonsillitis & 1.44 & 1.41 \\ 
		Common Cold & 1.15 & 1.00 \\ 
		Flu & 1.44 & 1.26 \\ 
		Cough & 1.76 & 1.63 \\ 
		\hline
		Average & 1.43 & 1.35 \\ 
		\hline
	\end{tabular}
\end{table}

\section{Conclusion}
\label{sec:conclusion}
We have created an integrated software system for monitoring disease outbreaks and generating explanations of detected events. The system provides tools for predicting the current levels of clinical case counts and for forecasting future levels of symptomatic social media activity. 

The DEFENDER architecture can be easily extended to handle new symptoms and geographical regions.
Adding new symptoms requires only the addition of relevant keywords. Extension to a new region requires the capture of geo-tagged tweets from the area, re-running the location clustering to generate a node network for the area and the addition of local news sources. Another main advantage is that the system can pick up signals from diseases that have not been selected in advance, due to the focus on symptoms. As long as the new disease shares symptoms with those already being tracked it will be picked up. The integrated situational awareness module allows the system to leverage the expressive power of social content combined with news media in order to provide causal explanations for detected outbreak events.
These design choices do however mean that diagnostic expertise is required to interpret symptom activity detected by the system. The use of the data driven location network also means that comparisons with clinical data from administrative regions requires data mapping and transformation.

The main contributions of this paper are fourfold. Firstly we have developed a 
novel technique to create a data driven location network from geo-tagged social 
media content. Secondly we have developed a generalised disease tracking 
approach which uses counts from multiple symptoms to predict current disease 
activity. Thirdly we attempt to forecast future symptom count levels by 
employing observed people movement from social media. Finally we have built 
these techniques, along with those developed in an earlier paper by the 
authors, into an integrated software system to aid public health officials 
working in syndromic surveillance.

We have evaluated each of the components of our system. In our earlier work 
\cite{thapen15} we tested the event detection and situational awareness 
algorithms. For event detection we examined 33 candidate alarms detected by the 
system. We manually assessed these alarms to identify their causes and where 
possible found external verification from clinical or news data that events had 
occurred. We compared these with those alarms tagged as genuine events by the 
system, producing an F1 score of 0.9362.
The situational awareness component was evaluated by determining the accuracy of the news linkage and tweet ranking algorithms. The news linkage, weighted towards precision, achieved
an F0.5 score of 0.79 on our example set of candidate alarms and produced no false positives at the optimum parameter level. The top ranked tweets fully matched our human-coded event summaries in 21 out of 26 cases that we examined.
Evaluation performed in this paper found that the social media data was
able to improve the tracking of diseases by 37 percent over a model that used
only previous case data. The forecasting of future symptom counts provided only
a moderate gain of 5.8 percent over a model using only previous count data.
An initial analysis of our location network using the PageRank algorithm revealed that nodes closer to the main 'trunk' of the UK running from London to Manchester gained in importance compared to their population, while seaside towns were less highly ranked due to their peripheral position in the network.

In future work we aim to develop a diagnostic model linking symptoms to specific diseases, incorporate additional signals into our situational awareness and event detection modules and expand the system to different regions of the world.

\section{Acknowledgements}

This research was carried out in cooperation with the UK Defence Science and 
Technology Laboratory. It was funded by the U.S. Department of Defense's 
Defense Threat
Reduction Agency (DTRA), through contract HDTRA1-12-D-0003-0010.
\newpage

\bibliographystyle{plain}
\bibliography{defender}

\end{document}